\documentclass[superscriptaddress]{emulateapj}
\usepackage{amsmath}

\shorttitle{Radiation Background Effects}

\begin{document}
\title{The Effects of the Ionizing Radiation Background on Galaxy Evolution}
\author{D. Clay Hambrick}
\author{Jeremiah P. Ostriker}
\affil{Princeton University Observatory, Princeton, NJ 08544}
\author{Thorsten Naab}
\author{Peter H. Johansson}
\affil{University Observatory Munich, Scheinerstr. 1, 81679
  Munich, Germany}

\email{dclayh@astro.princeton.edu}

\begin{abstract}
We find that the amount and nature of the assumed ionizing background
 can strongly affect galaxy formation and evolution.  Galaxy evolution
 simulations 
 typically incorporate an ultraviolet 
 background which falls off rapidly above $z=3$; e.g., that of
 \citet{hm96}.  However, this decline may be
 too steep to fit the WMAP constraints on electron
 scattering optical depth or observations of intermediate redshift
 ($z\sim 2-4$) Ly-alpha forest  transmission.
 As an  alternative, we present 
 simulations of the cosmological formation of individual galaxies
 with UV backgrounds that decline more slowly at high
 redshift: both a simple intensity rescaling and the background
 recently derived by \citet{fg09}, which softens the spectrum at higher
 redshifts.  We also test an
 approximation of 
 the X-ray background with a similar z-dependence. 
 We find for the test galaxies that 
 an increase in either the intensity or hardness of ionizing radiation
 generically 
 pushes star formation towards lower redshifts: although overall star
 formation in the simulation boxes is reduced by $10-25\%$, the
 galaxies show  a
 factor of $\sim2$ increase in the fraction of stars within a 30 kpc
 radius that are formed after
 $z=1$.  Other effects
 include  late gas inflows enhanced up
 to 30 times, stellar half-mass
 radii decreased by up to 30\%, central 
 velocity dispersions increased up to 40\%, and a strong reduction in
 substructure.  The magnitude of the effects depends on the
 environmental/accretion properties of the particular galaxy. 
\end{abstract}
\maketitle

\section{Introduction}
 The
importance of the ionizing background on the gas dynamics and hence
the star-formation history of galaxies has long been recognized
\citep{rees86,efst92}, and studied both numerically
\citep[e.g.][]{gnedin00} and 
semianalytically \citep[e.g.][]{bens02}.  (In this paper we take
``ionizing background'' to refer to both
the heating and H/He ionization effects of the 13.6 eV - 100 keV portion of
the external radiation field experienced by galaxies at various
epochs; i.e., the energies which can ionize H and other species but
have a large enough cross-section to significantly affect the gas
properties.)  However, although the hydrodynamic  
simulation of galaxies has become something 
of a cottage industry in recent years (see \citealt{mgk08} for a review with
  special focus on disk galaxies), relatively little attention has
been paid to the form of ionizing background which is used.  The default
option (seen recently in \citealt{gov07} and \citealt{scan08}, among others)
is generally a version of the UV background of \citet{hm96}.
Those authors modeled the QSO background and the effects of
reprocessing by Ly$\alpha$ forest clouds. They fit the redshift
dependence of the H photoionization rate with a generalized
Gaussian of the form $(1+z)^B \exp{[-(z-z_c)^2/S]}$; for their
parameters ($B=0.73$, $z_c=2.3$, $S=1.9$ in the original paper),
the intensity declines steeply above $z=3$ and the background is
negligible beyond $z=7$.  

There is, however, a history of simulations using ionizing backgrounds
with a less 
steep redshift dependence.  \citet{ns97} is one such example:
they performed $40^3$ SPH particle 
simulations resampled from a 
larger P$^3$M simulation, together with a UV background that was
constant at high redshift.  They found that the final amount of cooled
gas was reduced by up to half, with late-accreted gas preferentially
affected, compared to no background.   
Moreover, recent evidence such as the independent observations of the
optical depth to 
electron scattering \citep[WMAP: ][]{wmap} and the Lyman alpha
Gunn-Peterson trough 
\citep[SDSS: ][]{cfo03} have put increasingly tight constraints on the
reionization 
history of the universe.  In particular, \citet{fg08} examine the
Ly$\alpha$ effective optical depth using 86 quasar spectra, and find
an essentially flat ionization rate out to $z=4.2$ (see
Fig.~\ref{fig:spectra} and the discussion in \S2.1 below); these
authors have thus proposed a new UV background, described in
\citet{fg09}, which we investigate here. \citet{dww09}, using 1733
quasars from SDSS, similarly find a flat ionization rate for
$2<z<4.2$, albeit with a slightly different
normalization.  (\citet{hm01} themselves 
presented a UV background very similar to our New UV below, but that
model has seemingly failed to gain widespread acceptance.)  

Further, although the diffuse X-ray background is becoming increasingly
well constrained, at least in the local universe \citep{gilli07}, and
its unique role in heating the universe prior to full reionization
has been established \citep{venk01,rog05}, few
recent simulations \citep[for example]{ric08} have incorporated this
component.  The hard X-ray background is largely produced by QSOs and
lower-luminosity AGN.  Combining the the total background at $z=0$
with the observed X-ray spectra of individual sources and the redshift
dependence of AGN output permitted \citet{sos04} to estimate the $10^0
- 10^5$ keV background as a function of redshift.  This background has
a Compton temperature of $10^{7.3}$ K due to the peak in $\nu J_{\nu}$
around $30 - 50$ keV, which penetrates regions that are optically
thick to UV (although many simulations, including ours, treat the UV
as optically thin also) and can provide a considerable source of both
ionization 
and heating.  Indeed, \citet{me99} found that including an X-ray
background increases the equilibrium
temperature of the IGM by $\sim 20$\%.

Finally, since
the formation of massive ellipticals where star formation has been
effectively quenched since $z\sim1-2$ is still not well understood
and is the subject of ongoing studies
\citep[for a review see the introduction of][]{hop08}, 
a study to examine the extent to which results are sensitive
to the assumed ionizing radiation background is warranted.
(Naturally, further work could also be done with disk galaxies, Lyman-break
populations at higher redshift and others.)

The paper is organized as follows.  In \S\ref{sect:bkg}, we describe
the numerical 
methods and parameters of our simulations, in particular the
ionizing radiation backgrounds which we apply, comprising a recent
version of \citet{hm96}, a new rescaled version of the UV background 
which falls off much more slowly at high redshift, this new UV
background with an additional X-ray component, and the more realistic,
recently calculated UV background of
\citet{fg09}.   In
\S\ref{sect:results} we 
describe the results obtained from those simulations, in particular
the effects of the backgrounds on the gas properties, star formation and
stellar dynamics.  \S\ref{sect:disc}
is a discussion of the implications of these results when taken
collectively, and \S5 is conclusion.  

\section{Simulations and Backgrounds}\label{sect:bkg}
\subsection{Radiation Backgrounds}
\begin{figure}
\includegraphics[width=0.77\linewidth, angle=-90]{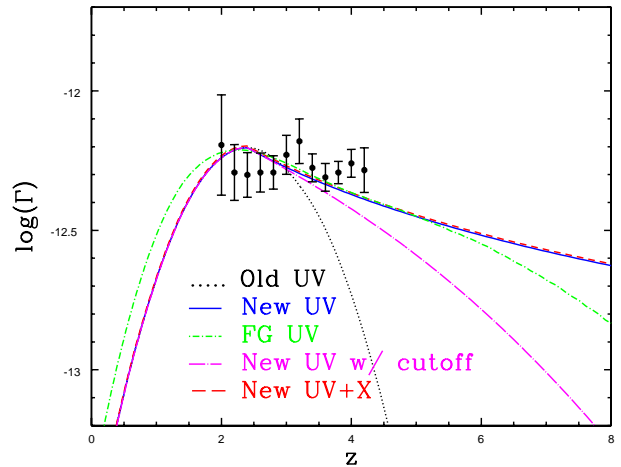}
\caption{Hydrogen photoionization rate $\Gamma$ (s$^{-1}$) vs.\ $z$ for the Old
  UV (black dotted line), New UV (blue solid line), New UV+X (red
  dashed line), FG UV (green short dash-dot line) and New UV with
  cutoff (magenta long dash-dot line) models.
  The data are from \citet{fg08}. Note that our 
  X-ray background contributes only slightly to H ionization,
  though it has significant heating and He ionization effects (see text).}
\label{fig:spectra}
\end{figure}

As a baseline, we use the \citet{hm96} UV background as used in e.g.\
\citet{naab07}; we call this the ``Old UV'' model.  To create an upper
bound on high-z UV, we keep the 
same assumed spectral shape but set the 
intensity to decline as $(1+z)^{-1}$ in physical units above the
\citeauthor{hm96} peak, 
which occurs at $z\approx 2.4$; we call
this the ``New UV'' model. For a more realistic case, we use the
background calculated by \citet{fg09}, which has a similar
z-dependence of intensity but whose spectrum softens
markedly for $z\gtrsim 3$ as the quasar contribution dies out; we call
this ``FG UV''.  The H ionization rates of these
three  models are compared 
in Fig.~\ref{fig:spectra}, along with the data of \citet{fg08}
mentioned above.   It is
apparent that New UV and FG UV provide a better fit to the
observations than does Old UV, in particular at $z>4$.

We also implement an X-ray background. This component uses the
spectral shape given in \citet{sos04}, which represents an average
quasar background considering both obscured and unobscured sources,
and is strongly peaked around 30 keV in $EF_E$.  The intensity
normalization comes from \citet{gilli07}, who modeled the AGN/QSO
X-ray background using both deep pencil-beam pointings and shallow
surveys.  This is
converted into heating and ionization rates using  Cloudy 
\citep[v07.02, last described in][]{cloudy}, which includes photo and
Compton heating as well as secondary ionizations; the
heating rates are then increased by a factor of 1.5 to better agree with the
more recent
model of \citet{socs05}. Heating
rates due to the UV and X-ray backgrounds are roughly equal for
virialized gas at $z=0$.   The redshift dependence of this background
is taken to be similar to New UV: intensity in physical units increases as
$(1+z)^3$ to $z=2$, and declines as $(1+z)^{-1}$ thereafter (which is
admittedly unrealistic; see below).
 We call this the ``New UV+X'' model.  Notice in Fig.~\ref{fig:spectra}
 that New UV+X has only a negligibly higher ionization rate than New
 UV: photons at keV energies and above deposit $\gtrsim 99\%$ of their
 energy as heat in a highly ionized medium through electron-electron
 collisions \citep{ss85}.  The X-ray background also contributes at
 the $\sim 30\%$ 
 level to the higher-energy HeII ionizations.  
 
As an additional motivation for the revised ionizing background, we
calculate a simple (homogeneous)
semianalytic model of 
reionization using Cloudy.  We create bins at successive epochs and
apply the corresponding background (based on our z-dependence
formulas) to gas of the corresponding (physical)
density (i.e., the mean present density scaled as $(1+z)^3$); Cloudy outputs
the electron-scattering and Gunn-Peterson optical 
depths ($\tau_{\text{es}}$ and $\tau_{\text{GP}}$, respectively). For
$\tau_{\text{es}}$ this method
is replaced above redshift 10 by an analytic integral of the electron
density (i.e., $\tau_\text{es}=\sigma_T \int n_e dl$ over the
appropriate cosmology, with 
$n_e$ given by the balance of photoionization with collisional
 recombination); FG UV is specified to have zero intensity
above $z=10$, 
so in that case this integral is $0$ also.  Note that removing the
assumption of homogeneity and 
including the effects of clumpiness would in general
decrease the effective recombination time \citep{mhr99}, thus
leading to a less-ionized universe and decreasing both
$\tau_{\text{es}}$ and the epoch of reionization; this strengthens our
conclusions below.  Moreover this calculation is optically thin and in
equilibrium, neither of which is the case for real reionization;
however our Gadget code also assumes optical thinness and equilibrium
(see below); so these results at least show what our simulations
would produce for $\tau_{\text{es}}$ and $\tau_{\text{GP}}$ and may be
compared to each other, even though
they are physically oversimplified and not directly comparable to
observations or more detailed calculations.   

The results are 
summarized in Table~\ref{tab:reion}, and 
compared with WMAP-5 \citep{wmap} and \citet{cfo03}
(CFO).  The Old UV model is significantly
($2.7\sigma$) low compared to WMAP in its prediction of
 $\tau_{\text{es}}$, while New 
UV and New UV+X are extremely high (6.7 and $9.2\sigma$,
respectively).  The evolution of the Gunn-Peterson
(Ly$\alpha$) optical depth $\tau_{\text{GP}}$ shows a qualitatively
similar pattern, 
with  the Old UV and New UV models comfortably bracketing the CFO
results.  FG UV agrees very well with both the WMAP $\tau_{\text{es}}$
and the CFO $\tau_{\text{GP}}$ values.  

The  values of $\tau_{\text{es}}$ for the New models may seem at first
too high to give
physically relevant results in a simulation; however, these  models exceed the
WMAP value only due to 
contributions from $z>10$, where it has been shown \citep{dijk04} that
the ionizing background has comparatively little effect.  (Again, FG UV
is specified to reach zero intensity at exactly $z=10$, and
matches the WMAP  $\tau_{\text{es}}$ value very well.) Even though
the gas may be highly ionized, its cooling time is still very short,
so the entropy and hence the dynamics are not much affected \citep{tw96}.
Therefore we expect our results not to differ so dramatically from a
WMAP-consistent model as the $\tau_{\text{es}}$ results would
suggest.  To test this hypothesis, we implement a high-z
cutoff for the UV of the form $[\exp(-((z-2.4)/4.7)^2), z>2.4]$,
analogous to \citet{hm96} with parameters
chosen to exactly match the WMAP-5 value of $\tau_{\text{es}}$, and
run a test simulation of Galaxy A to $z=0.5$ (this model is called
``New UV with cutoff'' in Table~\ref{tab:reion}).  We find this cutoff
causes essentially no
change in gas properties for all $z\leq5.2$, and only a small (1.5\%)
increase in total stellar mass at $z=0.5$ (by contrast the Old UV
simulation has 5\% more stellar mass than New UV at this redshift).
As we wish to ensure both that 
the new models 
represent a firm upper bound, and that any
differences from Old UV are large enough to be easily discernible, we
use the models without a cutoff in the rest of our analysis.  
These cutoff results also indicate that while New UV and FG UV have
slightly different redshift dependences for their intensities, those
differences will be negligible for our purposes compared to 
the differences in spectral shape of the two models.  In other words,
comparing the FG UV and New UV results will isolate the effect of the
spectral shape of the UV background, while comparing New UV to Old UV
will isolate the effect of its intensity (represented by e.g. H
photoionization rate) as a function of z.  In
fact, our results will show that for most of the properties we study,
increasing the intensity and increasing the spectral hardness 
 give the same qualitative effects.

Furthermore, with regard to the New UV+X model, we don't generically
expect UV and X-ray backgrounds to 
have the same z-dependence; in fact, since the X-rays primarily come
from quasars only 
rather than quasars and stars, and see a much smaller
optical depth, a more sudden dropoff in the X-ray background intensity
above the quasar luminosity function peak at $z\sim 2$ might be
expected.  However, again we are looking for an upper bound and for
any effects to be exaggerated for easy detection, and our overestimate
is relatively modest: at $z=3.9$, our New UV+X model gives a
mean-density IGM temperature $T_0$ of $\sim3.3\times10^4$ K, 50\%
higher than the observationally-fit value of $\sim2.2\times10^4$ K
\citep{zht01}.  

In sum, the FG UV model of the ionizing radiation background seems the
best of those considered in matching the observational constraints of
H ionization rate, electron-scattering optical depth, and
Gunn-Peterson optical depth.  In the following sections we will
explore the effect of this background on galaxy evolution, in
comparison with other backgrounds that are relatively too weak (Old
UV) and too strong (New UV, New UV+X).  

\begin{deluxetable}{lccc}
\tablecaption{Reionization with various backgrounds
\label{tab:reion}}
\tablehead{
\colhead{Name} & \colhead{$\tau_{\text{es}}$} &
\colhead{$z(\tau_{\text{GP}}=1)$} & \colhead{$z(\tau_{\text{GP}}=6)$}}
\startdata
\emph{WMAP-5} & $\mathit{0.087\pm 0.017}$ & --- & ---\\
\emph{CFO} &$\mathit{0.12\;\;\pm0.03\;\;}$ & $\mathit{4.21\pm 0.02}$ & $\mathit{5.95\pm 0.02}$ \\ 
Old UV & $0.040\pm0.002$ & $4.13\pm0.01$ & $4.78\pm 0.01$\\
New UV w/ cutoff & $0.087\pm 0.011$ & $4.09\pm 0.02$ & $5.46\pm0.02$\\
FG UV & $0.082\pm 0.011$ & $4.21\pm 0.02$ & $5.90\pm 0.02$\\
New UV & $0.202\pm0.011$ & $4.53\pm0.02$ & $6.28\pm 0.02$\\
New UV+X & $0.243\pm0.023$  &$4.77\pm 0.02$ & $6.49\pm 0.02$\enddata
\tablecomments{``CFO'' results are from \citet{cfo03}; ``WMAP-5'' results
  are from \citet{wmap}  These observational results are italicized.}
\end{deluxetable}

\subsection{Simulations}
We apply the different radiation fields to the SPH simulations
described in 
\citet{naab07}; see that paper  for full
details.  The code is based on GADGET-2, and the galaxies are
ellipticals selected from a (50 Mpc$/h$)$^3$ box with cosmological ($\Lambda$CDM)
initial conditions, and resimulated in a (10 Mpc$/h$)$^3$ box centered on
each galaxy, using high-resolution DM
and gas/star particles for the volume containing all particles which
are within 500 kpc of the cental galaxy at $z=0$ (this is roughly 1.5
Mpc$/h$ comoving).  Importantly, the simulation does not include optical
depth effects, in particular the self-shielding of dense star-forming
regions from the ionizing backgrounds, although those regions would be
optically thin to X-rays regardless.  Neither does it include any
feedback from supernovae or AGN; although those processes are
certainly important in many if not most galaxies, our object is show the
\emph{differential} effect of changes in the ionizing background on
the evolution of galaxies, independent of feedback
effects.  (However, it is interesting to note that recent work by
\citet{ps08} suggests that feedback and background effects may not be
independent, and may in fact amplify one another.)  Star formation is
performed at a fixed density threshold \citep[$\rho_\text{crit}=7\times
  10^{-26}$~g~cm$^{-3}$, or $n_{H,\text{crit}}=0.03$~cm$^{-3}$, as in][]{naab07}. 

 For easy comparison, we
use the same set of initial conditions that were designated
galaxies/halos A, C, and E 
in \citet{naab07}, and are so designated here as well.  All simulations
were performed with $100^3$ SPH particles (corresponding to a
gravitational softening length of 
0.25 kpc for the gas and star particles, and twice that for the dark
matter particles; gas and star particles have masses of the order
$10^6 M_{\odot}$).  As will be discussed below, galaxies A and C gave
the same qualitative results, while galaxy E was somewhat different,
due to its different merger and accretion history.  Since Galaxy A 
was the most well-studied in \citet{naab07}, we choose here to focus on
it, bringing in the other two galaxies where relevant.  (FG UV,
a later addition to the study, was run on galaxy A only.) 

Throughout the paper, all distances are physical except where noted;
the assumed cosmology is
$(\Omega_M,\Omega_{\Lambda},\Omega_b/\Omega_M,\sigma_8,h)=(0.3,0.7,0.2,0.86,0.65)$
 as in \citet{naab07}.

\section{Results}\label{sect:results}

\subsection{Gaseous effects}
We naturally expect an increase in the ionizing background to lead to
more efficient gas heating at high redshifts, but the consequences on
galaxy formation cannot easily be predicted.  

\begin{figure}
\includegraphics[width=\linewidth]{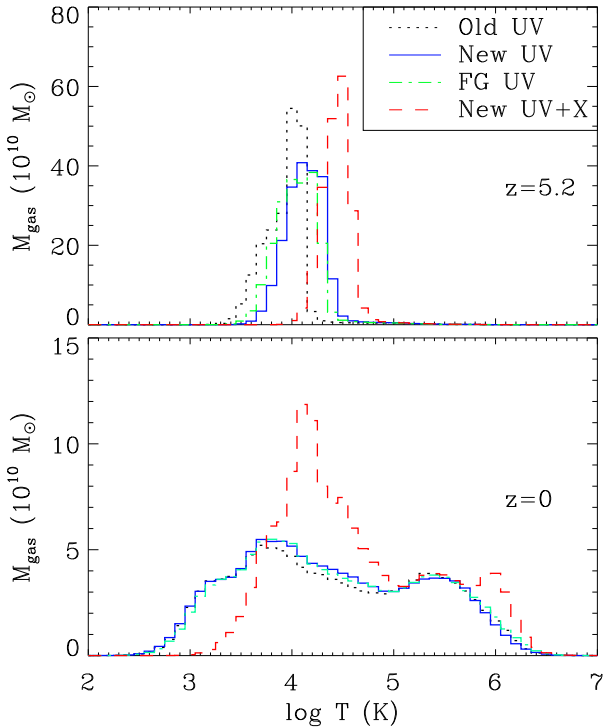}
\caption{Temperature distribution of gas at $z=5.2$ (top) and $z=0$
  (bottom) for simulation A ($r<2$Mpc comoving); 
  C and E have  similar results and are not shown.  At $z=5.2$ the Old UV,
  New UV and New UV+X
  models are clearly separated, while at $z=0$
  the addition of UV at early times has essentially no residual effect
  but the X-rays (which are present at all times) significantly heat the gas.}
\label{fig:Thist-all}
\end{figure}

Figure~\ref{fig:Thist-all} shows the temperature distribution of the gas
in the central 2 Mpc \emph{comoving} of the galaxy A
simulation (essentially the high resolution region of validity; the
simulations, being 
resampled from a larger box, are nonperiodic) with the three
backgrounds, at $z=5.2$  
(chosen to be where the Old UV background is nonzero but still substantially
lower than 
New UV / FG UV) and
$z=0$.  At $z=5.2$ the models are 
well separated; in fact the gas in the Old UV simulation has just
finished heating from a very cold (100K) state (to which it had cooled
by adiabatic expansion from its initial state at $z=24$; the
cooling function implemented in the code is primordial (H-He only) and
cuts 
off at $10^4$K), and has just finished H/HeI reionization.  As
expected, the New UV model, which has been entirely reionized
(including HeII) since
$z\simeq 8$,  
has significantly more hot gas, and including X-ray heating pushes most
of the gas up to temperatures significantly above $10^4$K.  The FG UV
model lies in between Old UV and New UV, which is expected since it
has a higher intensity than the former but a softer spectrum than the
latter.  FG UV has been H/HeI reionized since $z\simeq 8$ but has not
yet reionized HeII.

At $z=0$, however, the
Old UV, New UV, and FG UV models are essentially identical; increasing the
intensity or changing the spectrum of the ionizing background
at high redshift has no effect at the present (in other words, the
cooling time for all gas is shorter than 11 Gyr, since it has
forgotten the extra heat from $z>2.4$). This is consistent
with \citet{mbh06}, who found that the effects of a UV background
which is suddenly turned off
begin to dissipate after $\sim0.3$ of a Hubble time; 
since we are essentially turning off the \emph{extra} radiation at
$z=2.4$, we would expect a similar convergence.   Moreover,
\citet{hh03} found that well ($\gtrsim 2$ Gyr) after reionization, the
IGM equilibrium temperature approaches a value that depends only on
the spectral shape and not the intensity of the ionizing background,
so even if Old UV and New UV had different intensities all the way to
$z=0$ we wouldn't expect a significant difference.  

Adding X-rays (to
all epochs), on
the other hand, significantly heats the gas, especially the coldest
gas: the mean gas temperature rises 15\% from $1.3\times 10^5$ K to
$1.5\times 10^5$ K.  The addition of
X-rays also produces a 16\% larger total mass of gas due to reduced
star formation in small systems, as we will see in the next section.  
This larger reservoir of cool ($10^4 - 10^5$ K) gas (cf. the lower
panel of Fig.~\ref{fig:Thist-all}) helps to
prolong and enhance the epoch of star formation in massive
galaxies. The warm-hot gas mass (WHIM; 
$10^{4.5}<T<10^7$ K) is 30\% larger in 
the X-ray case than in the cases without X-rays.

\begin{figure}
\includegraphics[width=\linewidth]{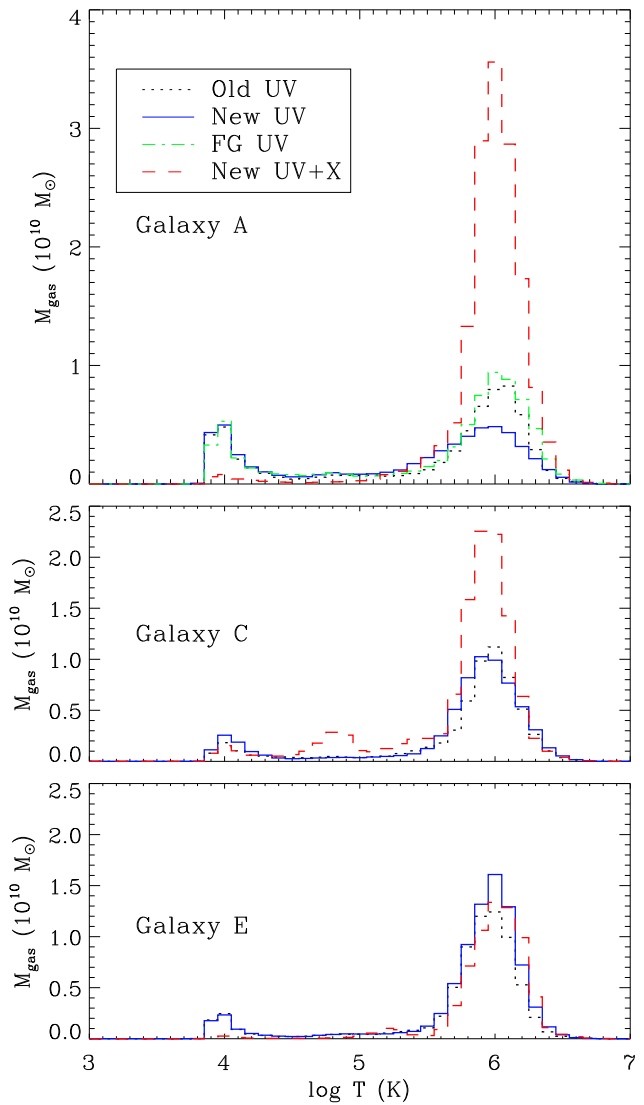}
\caption{The same as Fig.~\ref{fig:Thist-all} ($r<2$Mpc; $z=0$), but for gas with
  $\rho>200\overline{\rho}$, for galaxies A (top), C
  (middle), and E (bottom).  Adding more UV at early times has little
  effect, while adding the X-rays removes the
  cold-dense gas while substantially increasing the hot dense gas in
  two of three cases. }
\label{fig:Thist-high}
\end{figure}

Of course, we are especially interested in the gas which has collapsed
and virialized in dense halos.  Figure~\ref{fig:Thist-high} shows the
temperature spectrum of high-density gas 
($\rho > 200\overline{\rho}$, where $\overline{\rho}$ is the mean
baryonic density of the universe) at $z=0$ for the four backgrounds and
the three galaxies.  (We note again that these simulations include no
optical depth effects and therefore overestimate the UV flux that
virialized regions see.)  We see that adding early UV does not affect
the amount of cold ($10^4$K) dense gas, and has uncertain effect on
the hot ($10^6$K) dense 
gas, increasing it slightly in galaxy C, making negligible change in galaxy E
and decreasing it somewhat in galaxy A (although not for FG UV).  Adding an
X-ray background 
substantially increases the hot dense gas for galaxies A and C, while
having negligible effect on galaxy E.  

We explain these differences by reference to the merger histories of
the three galaxies.  As explored in \citet{naab07}, galaxy A has a
merger of mass ratio $6.5:1$ at $z\approx0.6$ (6 Gyr ago), galaxy C
has a merger of mass ratio $3.5:1$ at $z\approx 0.8$ (7 Gyr ago),
while galaxy E has no significant merger events after an equal-mass
merger at $z\approx 1.5$ (10 Gyr ago).  Since the gas-to-star ratio is
a strongly declining function of halo mass, especially when there is
significant ionizing radiation to keep the low-density gas in small
halos hot, we expect in New UV+X for the accretion of
smaller halos 
at later times to add more hot gas compared to accreting
larger halos at earlier times, and the earlier the gas is added to the
dense central galaxy, the more of it can cool and form stars in situ.  On the
other hand, the mergers in the 
New UV (no X) case involve large amounts of colder gas (see next paragraph),
which mingles with the existing gas and 
forms stars quickly, thus paradoxically resulting in less hot gas at the
present for Halo A.

\begin{figure}
\includegraphics[width=\linewidth]{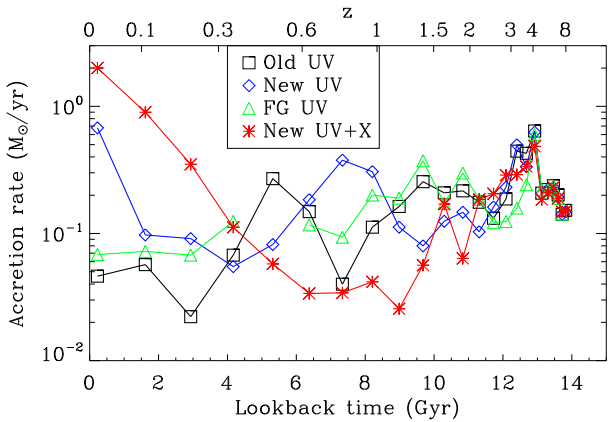}
\caption{Accretion rates of gas  onto the
  central 10kpc (physical), 
  for galaxy A.  
  The New UV and New UV+X models have increased accretion at
  late times (LBT$<4$Gyr). (Missing data points indicate zero
  accretion or net outflow.) }
\label{fig:cgasacc-A100}
\end{figure}

Figure \ref{fig:cgasacc-A100} shows the accretion rate of gas
  onto the 
  central physical 10kpc of galaxy A; galaxies C and E had qualitatively
  similar results.  We see immediately that  New UV and New UV+X have
  significantly higher accretion compared to Old UV, especially in the
  last 4Gyr ($z<0.3$) (and in the no X-ray case, the peak at $\sim 8$
  Gyr ago, corresponding 
  to the merger event mentioned above); FG UV has slightly enhanced
  accretion.    This is easily understood as
  gas that was kept 
  hot at early times by the harder early background of New UV and New
  UV+X \citep[as per][]{hh03} finally cooling and
  flowing inward; in the model with lower UV, much of this gas would
  have formed small stellar systems and been unavailable for late
  inflows.  At $z=0$, the inflowing gas in the New UV model has a mean
  temperature of $4\times 10^5$ K, while for New UV+X the mean
  temperature is $3\times 10^6$ K.

\subsection{Stellar effects}
\begin{figure}
\includegraphics[width=\linewidth]{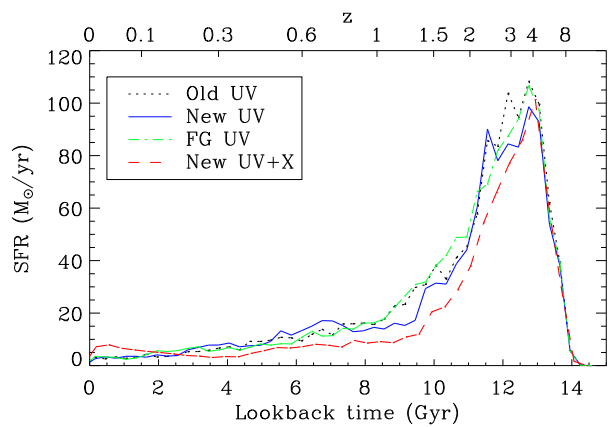}
\caption{Star formation history of the central 2 Mpc of the Halo A
   simulation, for the 
  four backgrounds.  The high-z peak is suppressed as radiation
   increases (meaning as we move from
Old UV to FG UV to New UV to New UV+X, increasing the intensity and
   hardness of the ionizing background).  }
\label{fig:sfr-box-A100}
\end{figure}

The effects of the different background models on star formation fits
well with what we saw in the gas.  Figure~\ref{fig:sfr-box-A100} shows
the star-formation history of the central 2 Mpc of simulation A (that
is, the formation history of stars in the central 2 Mpc at $z=0$).  As we
would expect, the early 
star star formation peak at $z\sim 4$ is increasingly suppressed as we
increase the intensity and hardness of the ionizing background, from
Old UV to FG UV to New UV to New UV+X (and this carries 
forth to the 
total stellar mass at the 
present, which is decreased by 1.4\% for FG UV, 7\% for New UV, and
 24\% for New UV+X 
compared to Old UV; see Table~\ref{tab:stars}).
However, the increased radiation above $z=2.4$ in the New UV model has
little effect on global star formation below that redshift.  In
contrast, New UV+X suppresses star formation until the last 2 Gyr,
where there is a modest bump from the inflow of late-cooling gas.   

\begin{deluxetable}{llccc}
\tablecaption{Stellar mass results
\label{tab:stars}}
\tablehead{
\colhead{IC Name}&\colhead{Background} & \colhead{$M_R(5\text{kpc})$}& \colhead{$M_R(30\text{kpc})$}& \colhead{$M_R(2\text{Mpc})$}}
\startdata
A & Old UV & 9.28 & 14.00 & 37.45\\
A & FG UV & 10.09 & 14.14 & 36.92\\
A & New UV & 10.23 & 14.72 & 34.95\\
A & New UV+X & 9.91 & 13.91 & 28.52\\ 
C & Old UV & 8.55  & 14.12  &36.34 \\
C & New UV & 9.28  & 16.18  &35.58 \\
C & New UV+X & 10.81  & 16.69  & 29.58 \\ 
E & Old UV & 8.24  & 12.43 & 27.98\\
E & New UV & 8.78 & 16.71 & 27.75\\
E & New UV+X & 12.66 & 13.78 & 25.29 
\enddata
\tablecomments{Results are the mass of stars within the
  specified radius of 
  the principal halo at $z=0$.  Masses are $10^{10}M_\odot$.}
\end{deluxetable}

\begin{figure}
\includegraphics[width=\linewidth]{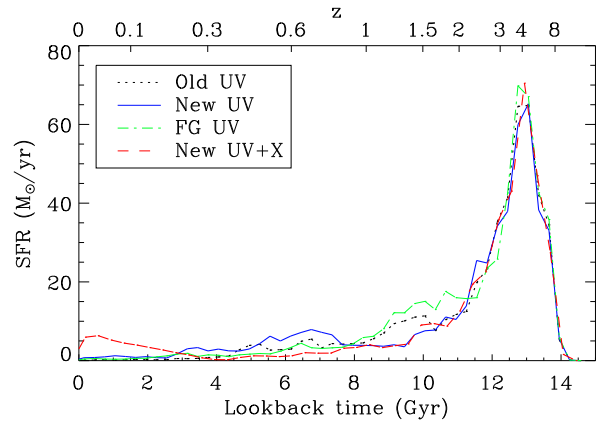}
\caption{The same as Fig.~\ref{fig:sfr-box-A100}, but only considering
  star particles which are within 30 kpc of the central halo at $z=0$.
  There is increased star formation at low redshift ($z<1$) as
  radiation increases: the mean (mass-weighted) age of these stars
  decreases from 11.4Gyr (Old UV) to 10.9Gyr (New UV+X).}
\label{fig:sfr-gal-A100}
\end{figure}

We also study the star-formation effects in a single, dense region:
viz., the central 30 kpc of galaxy A, presented in
Fig.~\ref{fig:sfr-gal-A100}.  Here there is little effect on the
initial peak, which we expect since this region should have a short
cooling time due to its overdensity.  However, star formation below
$z=1$ is enhanced: 19\% of the stellar mass within 30kpc is formed after
$z=1$ in the New UV case, compared to 11\% for Old UV (and 10\% for FG UV).
The bump in New UV centered at $\sim$7~Gyr ago matches well with
the merger-event peak in gas accretion we saw centered at
$\sim$8~Gyr ago in  
Fig~\ref{fig:cgasacc-A100}.  This results in the New UV version of
galaxy A
actually having 5\% \emph{more} stars at the present within the 30 kpc radius
compared to Old UV; FG UV has only a 1\% increase (again see
Table~\ref{tab:stars}).   New UV+X has a very late burst of star formation 
 0.5 Gyr ago which is
due to a the infall of warm gas that cools in situ, but this is not
enough to make up for the earlier deficit, and it has slightly less
stellar mass within 30 kpc than Old UV; 15\% of the stellar mass is
formed after $z=1$.  

Finally, we convert these star-formation histories into
present luminosities via the {\sc galaxev} code of \citet{bc03},
assuming a Salpeter IMF and solar metallicity;
bolometric magnitude and color results for Galaxy A are presented in
Table~\ref{tab:obs}.  No 
dust extinction effects were included.  In brief, we see that compared
to Old UV, New UV has only a slight increase in bolometric luminosity
(0.13 mag) and blueness (0.09 mag), and FG UV has even less change,
while New UV+X is 1.3 mag or 
3.25 times brighter, and 0.28 mag bluer in U-B, thanks to its late
burst of star formation.  

\begin{deluxetable}{lccccc}
\tablecaption{Observational Characteristics, Galaxy A
\label{tab:obs}}
\tablehead{
\colhead{Model Name} & \colhead{Bol.\ Mag.} & \colhead{B Mag.}
& \colhead{R Mag.} & \colhead{U-B} & \colhead{B-V}}
\startdata
Old UV & -21.91 & -20.16 & -21.72  & 0.59  & 0.93\\
FG UV & -22.00 & -20.30 & -21.82 & 0.55 & 0.91\\
New UV & -22.04 & -20.39 & -21.85 & 0.50 & 0.87\\
New UV+X & -23.19 & -21.96 & -23.14 & 0.31 & 0.69 
\enddata
\tablecomments{Results are for a radius of 30 kpc at $z=0$.}
\end{deluxetable}



These results, like those for the gas, are dependent on the
environment of the galaxy.  For example, with the Galaxy E simulations
the New UV+X 
model has an 10\% decrease in total stellar mass compared to Old
UV, but has 34\% \emph{more} stellar mass in the central 30kpc from
enhanced gas accretion.  (New UV with no X-rays produces 1\% less
total stellar mass and an 11\%
increase within 30 kpc.)  There is also a correspondingly stronger
effect on the mean stellar age for galaxy E: the fraction of stellar 
mass within
30kpc formed after $z=1$ increases from 6\% for Old UV to 35\% with 
New UV+X.  See \S\ref{sect:disc} for 
discussion of these differences.

\subsection{Dynamical effects}

\begin{figure}
\includegraphics[width=\linewidth]{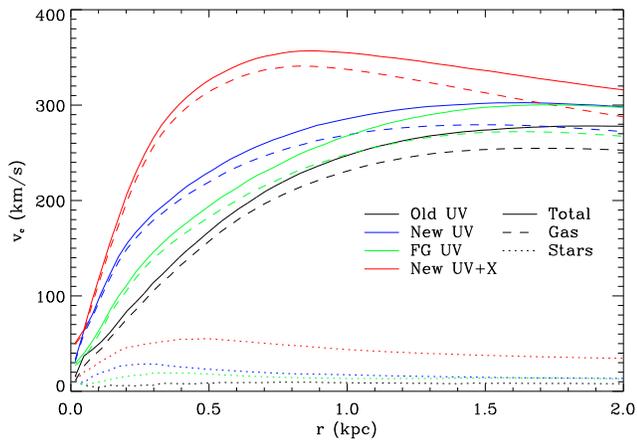}
\caption{Circular speed for stars (dashed line), gas (dotted line), and total
  (solid line; also includes dark matter), for Halo A with the four
  backgrounds, $z=0$. Colors for the four background models are 
  as before.  The peak
  value and inner slope increase with increasing radiation.}
\label{fig:vcirc-A100-z0}
\end{figure}

In addition to affecting the mass and hydrodynamic properties of the
stars and gas in the simulations, the radiation background affects
their arrangement; i.e., the dynamical properties of the galaxy.  The
first column of Table~\ref{tab:stars} shows a significant increase in the
stellar mass 
within a 5 kpc radius as one moves from the Old UV to the
New UV to
the New UV+X case for all three galaxies (and FG UV again lies between
Old UV and New UV): the mass increase from Old UV to
New UV is an average of $\sim$10\%, and from New UV to New UV+X
averages $\sim$20\%, although this includes a slight decrease for
Galaxy A.  This increase is natural considering
the enhanced gas flows we saw in 
Fig.~\ref{fig:cgasacc-A100}: once that gas reaches the center it must
cool and form stars there.  
Figure~\ref{fig:vcirc-A100-z0} shows the circular speed $v_c^2=GM(r)/r$ for the
innermost 2 kpc of galaxy A at $z=0$.  We see that the stars become more
centrally concentrated as one moves up in background
intensity or hardness (from Old UV to FG UV to New UV to New UV+X); this is
a general feature for $z<2.5$.  Residual gas also increases in the
same way.

\begin{figure}
\includegraphics[width=\linewidth]{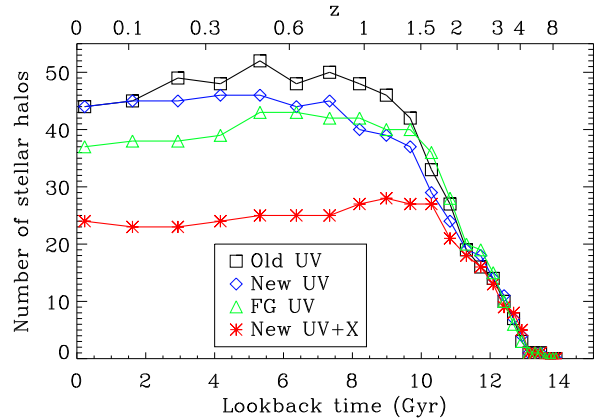}
\caption{ Number of discrete stellar systems
with masses $\geq 3\times 10^8 M_\odot$ identified by HOP
for the galaxy A simulation (within 2 Mpc comoving) over time.  New UV
suppresses small halos somewhat for $0.1<z<1.5$, while New UV+X
does so strongly for all $z<1.5$; FG UV shows both the suppression of halo
formation ($1.5>z>0.75$) we see in New UV and the late mergers
($z<0.5$) of Old UV, resulting in fewer systems than either (though
still many more than New UV+X. }  

\label{fig:haloint-A100}
\end{figure}

Figure~\ref{fig:haloint-A100} shows the number of discrete (i.e.,
non-substructure)  stellar systems
with masses $\geq 3\times 10^8 M_\odot$ identified from the star
particles with the HOP
halo-finding algorithm \citep{hop} 
for the galaxy A simulation (within 2 Mpc comoving) over time.  Many
past simulations 
\citep{efst92,qke96,tw96} have shown that photoionization can prevent the
formation of smaller stellar systems, and indeed we find that the number
of total stellar systems decreases from Old UV to New UV to New UV+X, although
we find the effect of increased high-z UV on the number of independent
halos doesn't persist
past $z=0.5$, where 
 merger events consume several small halos that have formed in
the Old UV case. Interestingly, FG UV shows fewer small halos than
New UV: we see the suppression of halo formation ($1.5>z>0.75$)
combined with late mergers ($z<0.5$).  In all cases 
star formation in smaller halos is
preferentially suppressed.  A future paper (Hambrick \& Ostriker
2009, in prep.) will deal in more detail with the effect of radiation
backgrounds on the mass spectrum of galaxies.

\begin{figure}
\includegraphics[width=\linewidth]{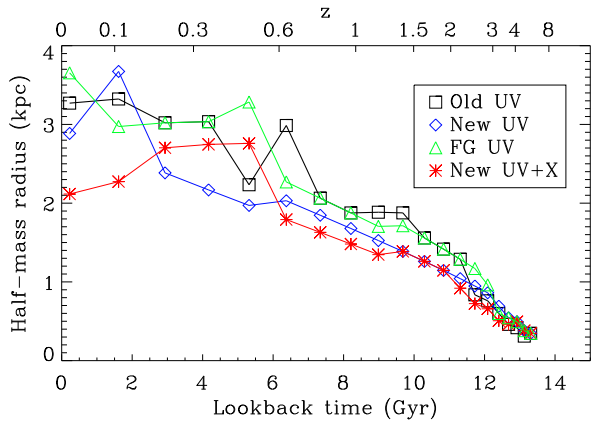}
\caption{Half-mass radii of stars within 30 kpc for Halo A.
  The increased radiation models show consistently smaller 
  radii.  (The large jumps are the result of merger events.)}
\label{fig:hmr-A100}
\end{figure}

Figure~\ref{fig:hmr-A100} shows the half-mass radius for stars within
30kpc for the three radiation models with galaxy A.  The major effect
is secular growth from $z=8$ to the present, combined with the
stochastic effects of major mergers.  \citet{naab09}
found in a detailed examination of galaxy A (with the 
Old UV background, and including SN feedback) that minor mergers are
primarily responsible for the 
factor of 
$\gtrsim 3$ increase in half-mass radius from $z=3$ to 0. The
radiation background does have an effect, however: New UV and
New UV+X show 
substantially ($\gtrsim25\%$) smaller radii compared to Old UV at all
$z<2.5$, modulo the 
intermittent merger effects; FG UV has negligible difference from Old
UV, and in fact has a slightly larger radius at $z=0$.    

For $z\gtrsim 3$ the picture looks somewhat different:
specifically, in our snapshot at $z=2.9$, the New UV model for galaxy
A has a 13\%, and FG UV a 27\%
\emph{larger} stellar half-mass radius than Old UV, (although New UV+X
is smaller than Old UV by 13\%).  Further, New UV and New
UV+X both show a 5\% reduction in peak circular speed compared to Old
UV at that redshift. 
\citet{cen08} found, 
using a modified \citealt{hm96} background similar to our ``New UV
with cutoff'', that simulated galaxies
have half-light radii that are too small and peak circular speeds that
are too large compared to observed galaxies at redshift 3.  However,
these authors also used a more stringent star-formation criterion:
when they repeated their simulations using the same density threshold
as in this work, their results at $z=3$ agreed with obesrvations (Ryan
Joung, priv. comm.).

We speculate that the increase in central concentration with
increasing radiation is enhanced by reduced substructure in the galaxy
(because small stellar halos are suppressed) and correspondingly less
dynamical friction.  To test this hypothesis, we have constructed a
statistic to measure the substructure or
``lumpiness'' that we designate $\sigma_{\text{UM}}(r)$, which is
created by an unsharp-mask-like procedure:
\begin{align*}
\sigma^2_{\text{UM}}(r) =&
\frac{1}{N}\sum^N_{i:\genfrac{}{}{0pt}{}{R_e/5<R(i)<2R_e}{U(i)>0}}
\mspace{-40mu}U(i)^2\\
U(i)=&\frac{\rho(i)-(\rho\circ G(r))(i)}
   {(\rho\circ G(r))(i)},
\end{align*}
where $R_e$ is the effective (half-mass) radius of the galaxy, $\rho$
is the stellar density field (created by cloud-in-cell mapping of the
star particles to a grid of resolution 0.38 kpc, or twice the gas
softening length), and $G(r)$ is a gaussian of width $r$.  
That is, we
create a mask by smoothing the stellar density field of the galaxy
with a fixed-width gaussian 
kernel, then sum $((\text{data}-\text{mask})/\text{mask})^2$ over all
pixels in the grid where it is positive 
(i.e. overdense regions) except the central peak, and normalize to the
size of the galaxy.  We 
find that the Old UV model indeed has far more substructure than the
two new models, at least for halos A and C; Table~\ref{tab:substruct} shows
$\sigma_{\text{UM}}(4\text{kpc})$ for the three backgrounds and ICs.
Halo E, as remarked on in \citet{naab07}, is composed mainly of stars
which formed in-situ rather than the accretion of smaller stellar systems,
and therefore we expect its substructure to be much less affected by
the ionizing background, which is what we find.  

At first glance, Fig.~\ref{fig:haloint-A100} and
Table~\ref{tab:substruct} may seem to be incompatible, since the two
models without X-rays are nearly the same in the former and vastly
different in the latter. However, if we
increase the outer radius in the definition of $\sigma_{\text{UM}}$ from two
to three effective radii (roughly 
from 20 to 30 kpc for Galaxy A), we find that
$\sigma_{\text{UM}}(4\text{kpc})$  
becomes $(125.7, 145.8, 31.5)$ for Old UV, New UV, and New UV+X
respectively, a result much more in accordance
with
Fig.~\ref{fig:haloint-A100}.  That is, the New UV simulation has
numerous small 
halos, but they are at larger radii from the central galaxy than in
the Old UV case.  On the other hand, Old UV and New UV have the same
number of independent halos at the present, as seen
Fig.~\ref{fig:haloint-A100}, but Old UV has significantly more
substructure.  This means that Old UV formed more small halos
initially, but they were accreted onto the central galaxy by
$z\simeq0.5$, enhancing its substructure.  

\begin{figure*}
\centerline{\includegraphics*{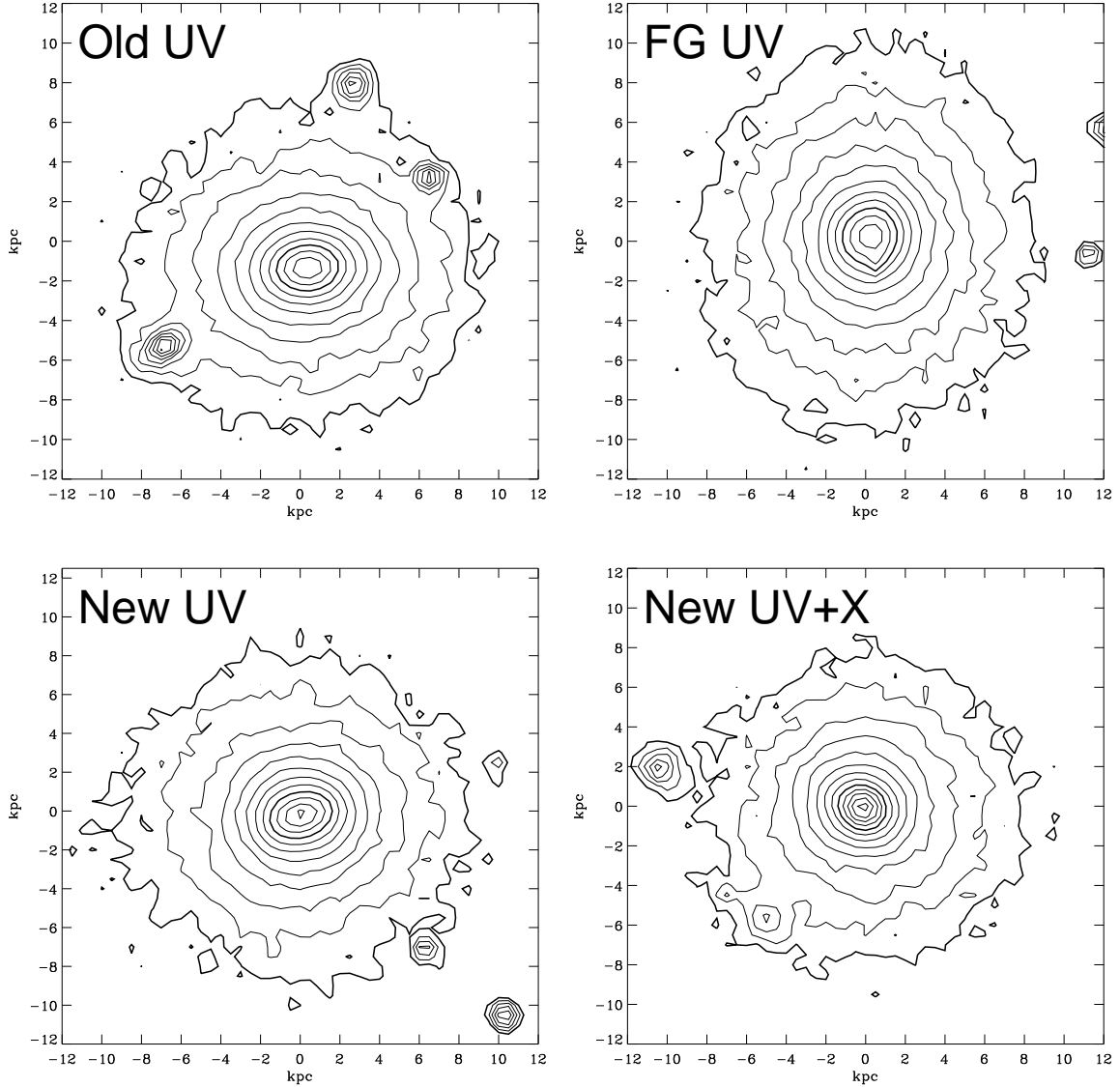}}
\caption{Contours of stellar density in the central 20kpc of galaxy A
  at $z=0$ for the Old 
  UV, FG UV, New UV,  and New UV+X  models.  
 The scales are kpc; the contours are
  quarter-decade steps in stellar density, with the inner bold contour
  corresponding to $3M_\odot/\text{pc}^3$.  The reduction in size and
  substructure (i.e. the number and size of subhalos) for New UV models are
  both visible, as is the increased central peakiness.} 
\label{fig:starcont}
\end{figure*}

Figure~\ref{fig:starcont} shows contours of stellar density for
the central 20kpc of galaxy A with the four
background models.  The smaller size of New UV+X is readily apparent,
as is the reduction in the number and mass of subhalos.  One can also
see the sharp central peak in New UV and New UV+X which is absent in Old UV. 

\begin{deluxetable}{lccc}
\tablecaption{Substructure Measure
\label{tab:substruct}}
\tablehead{
\colhead{Model Name} & \colhead{A} &\colhead{C} & \colhead{E}}
\startdata
Old UV & 114.6 & 98.9 & 113.9\\
FG UV & 19.1 & - & - \\
New UV & 14.2 & 13.7 & 129.8\\
New UV+X & 31.5 & 17.8 & 70.6\\
\enddata
\tablecomments{Data are $\sigma_{\text{UM}}(4\text{kpc})$; see text
  for definition.}
\end{deluxetable}


\section{Discussion}\label{sect:disc}
Taken collectively, our results suggest a picture of the effects of an
increased early UV radiation background along the following lines.  At
early times ($z\gtrsim 2$), the more intense/harder radiation and earlier
reionization make the gas
hotter, especially the less-dense gas, which cannot cool as
effectively (Fig.~\ref{fig:Thist-all}).  That is to say, the
processes of H and He reionization create large injections of heat, so
even though the gas begins to cool one reionization is complete, an
earlier reionization means the gas spends less time in the cold,
neutral state it has been in since recombination.  At very early times
($z>4$) the gas 
has not yet equilibrated, and so the more intense but softer FG UV
background gives a higher temperature than Old UV, but by $z=3.2$ the
situation has reversed, consistent with \citet{hh03} .  At late times, this
means that 
the primary
galaxy has less substructure, since there are fewer small stellar systems to
accrete; it contains more gas in its central regions
(Fig.~\ref{fig:cgasacc-A100}), and hence more late 
in-situ star formation (Fig.~\ref{fig:sfr-gal-A100}), since that gas would
otherwise have formed stars before falling in; it is still smaller but
now more tightly bound due to gaining less energy from
dynamical friction \citep{naab09} and more central star formation from
infalling cold 
streams.  In the terms of the \citet{bdn07} picture, the cold gas
inflows persist longer when early star formation is suppressed by
ionizing radiation.  

We can distinguish the effects of intensity and spectral shape by
comparing New UV and FG UV, since they have roughly the same intensity
(H photoionization rate, etc.) from $2<z<8$ but FG UV has a
significantly softer spectrum due to the rapid falloff of the quasar
contribution.   We find that FG UV in general gives results
intermediate between Old UV and New UV, for gas temperature, rate of
gas accretion by the central galaxy,
star-formation rate, and concentration of stellar mass; in other
words, an increase in the intensity of early radiation and an increase
in the hardness of the background spectrum produce qualitatively the
same results in most of properties we study here.  One exception
is in the total number of independent stellar systems: FG UV has
somewhat fewer than both Old and New UV (though still far more than
New UV+X).  

Continuing the investigation of spectrum, the effects of X-ray as
opposed to UV radiation are equally 
interesting.  Since adding the X-rays increases radiation levels
at all $z$, not just $z>2.4$, the gas at the present is substantially
hotter, and star formation is pushed to even later times, in agreement
with \citet{jno09}, who found that the accretion of small stellar
clumps (minor mergers) was sufficient to suppress star formation after
$z\approx 1$: in the New UV+X case the small stellar clumps are
themselves suppressed, so the central galaxy continues star formation
to the present.  There is less star formation overall, however, and
therefore more dense (hot) gas. 
Substructure is further reduced and compactness increased \citep[again
  agreeing with][]{naab09}.

These effects are, however, dependent on the accretion
history of the galaxy in question.  \citet{naab07} described two
distinct mechanisms by which galaxies assemble their stellar mass:
the accretion of existing stellar systems, and the in situ formation of stars
from inflowing gas.  In our sample, the former mechanism is dominant
for galaxies A and C, and the latter for galaxy E.  The increase in
ionizing radiation creates a bias toward the gas-accretion mechanism
by suppressing star formation in small halos.   Thus in the galaxies
where stellar accretion was important in the low-radiation (Old UV)
case, A and C, we find that the increase of in situ star formation is
more or less balanced by a decrease in accreted stellar mass, so the
total stellar mass in the galaxy at the present is not strongly
affected, although the mean stellar age and hence the luminosity and
color are.    However, in galaxies where stellar accretion is
not important (like galaxy E), there is no mass loss from that source,
so extra surrounding gas falls in and leads to a higher total galaxy
stellar mass.  That is, gas which in the Old UV case formed stars in
small satellite halos gets heated up at early times, then at late
times ($z<1$),
flows in and forms stars in the large central galaxy.

\section{Conclusions}  
As we develop accurate, detailed modeling for galaxy formation
based on realistic cosmological initial conditions, we are learning
just how sensitive the results are to the physical input parameters.
One important parameter is the ionizing background radiation field.
The commonly 
accepted \citet{hm96} model, while it has been of great utility, fails
to produce a good match to the most relevant observational
constraints: first, the ionization rate as determined by Gunn-Peterson
observations of high-redshift quasars, and second, the electron
scattering optical depth as determined by WMAP.  

We have considered
new models for the ionizing background: a simple rescaling of
\citet{hm96}  both with and without a 
significant X-ray component,  intended to represent an upper bound of
what is possible, and the recent results of \citet{fg09}, representing
a reasonable middle ground.  We have performed a state-of-the-art set of
cosmological simulations to assess the sensitivity of the results to
the assumed ionizing background.  

 We find that the gas properties at late times are much more affected
by the X-ray component than by early UV enhancement, with the result
that there is as much as a 30\% increase in the WHIM component
($T>10^{4.5}$K), and a four-fold increas in the hot-dense component
($T\approx 10^6$ K; $\rho > 200 \bar{\rho}$) when X-ray heating is
present.  However even our rescaled 
background without X-rays (New UV) reduces the
formation of stars in small systems and allow cold 
flows to persist to later times, markedly increasing the amount of
late-time ($z<1$) star formation in massive galaxies.  
Correspondingly there is less substructure in the massive systems due
to reduced accretion of smaller stellar systems and consequently
less gravitational heating.  The systems do still grow in size due to
accretion of satellite systems but the effect is less pronounced with
when X-ray heating.  
Despite the absence of feedback, and at a
resolution of $100^3$ particles (with gas softening length of 0.1625/h
kpc), all simulations have a star-formation 
rate of less than $1
M_{\odot}$/yr 
at $z=0$ with the exception of the model with X-rays, in which cold
flows persist to the present.
Finally, the mass function for small-mass systems is somewhat reduced
by extra early UV and substantially so with the addition of X-rays;
this is a topic which will be explored in greater detail in a future
paper.  

The background of \citet{fg09} (FG UV), in contrast with New UV, matches both 
observed ionization rate and electron-scattering optical depth, and
reduces the substructure in massive systems \emph{without} creating
galaxies that have too much gas, too many new stars, and too small a
size at late times.  Therefore we recommend its adoption over all the
models studied here for future galaxy simulations.

JPO was supported by NSF grant AST 07-07505 and NASA grant
NNX08AH31G. DCH thanks Luca Ciotti and Andrei 
Mesinger for helpful advice.   TN and PHJ acknowledge support by the
DFG cluster of excellence `Origin and Structure of the Universe'; TN
thanks Shy Genel for an interesting discussion.   The
authors are especially grateful to 
the referee for many good suggestions and particularly for bringing
the FG UV background to our attention.

\end{document}